\documentstyle[12pt]{article}

\def\bq{\begin{quotation}}
\def\eq{\end{quotation}}

\def\fnote#1#2 {\begingroup \def \thefootnote {#1}
\footnote{#2}\addtocounter{footnote}{-1}\endgroup}

\newcommand{\myname} {\vspace{0.5in}
                                  \begin{center}Zhu Yang\\
                                  \vspace{0.2in}
                                  Department of Physics and Astronomy\\
                                  University of Rochester\\
                                  Rochester, NY 14627\\
                                  \vspace{0.5in}
                                 Abstract\\
                                  \vspace{0.2in}
                                \end{center}}

\newcommand{\pagenumber}{\pagestyle{plain}\setcounter{page}{1}}

\def\a{\alpha} 
\def\b{\beta} 
\def\d{\delta}

\def\l{\lambda} \def\L{\Lambda}
\def\m{\mu} 
\def\n{\nu}

\def\s{\sigma} 
\def\t{\tau}

\def\raisenot{\raise .5mm\hbox{/}}
\newcommand{\notpa}{\hbox{{$\partial$}\kern-.54em\hbox{\raisenot}}}
\def\notp{\ \hbox{{$p$}\kern-.43em\hbox{/}}}
\def\notq{\ \hbox{{$q$}\kern-.47em\hbox{/}}}
\def\notk{\ \hbox{{$k$}\kern-.47em\hbox{/}}}
\def\notA{\ \hbox{{$A$}\kern-.47em\hbox{/}}}
\def\nota{\ \hbox{{$a$}\kern-.47em\hbox{/}}}
\def\notb{\ \hbox{{$b$}\kern-.47em\hbox{/}}}

\begin {document}
%\QQc
\baselineskip=24pt

\pagestyle{empty}

\begin{flushright}UR-1336\\
ER-13065-678\\
Oct. 1991
\end{flushright}
\vspace{0.5in}
\begin{center}
{\Large
A Remark on Conformal Anomaly and Extrinsic Geometry of Random Surfaces}
\end{center}

\myname

In low dimensions, conformal anomaly has profound influence on the
critical behavior of random surfaces with extrinsic curvature rigidity $1/\a$.
We illustrate this by making a small $D$ expansion of rigid random
surfaces, where a non-trivial infra-red fixed point is shown to exist.
We speculate on the renormalization group flow
diagram in the $(\a,D)$ plane. We argue that  the qualitative
behavior of numerical simulations in $D=3, 4$ could be
understood on the basis of the phase diagram.

\newpage
\pagenumber

Random surfaces are being studied in many branches of physics
\cite{gsw,amp,npw}. It is thus
important to understand their phase structure better. Typically one
starts from Nambu-Goto action. Since it contains smallest number of derivative,
it is supposed to be the most relevant term in infrared region.
The Nambu-Goto model describes physics well as long as one treats it as a
effective theory with a cutoff.
The continuum limit, corresponding to
a collapsed or branching phase,  does not seem to be
interesting. Such a collapsed phase
is avoided in condensed mattering physics by adding non-local
interactions \cite{npw}.
There are other situations where the surface tension becomes small
(by fine tuning), and
higher derivative terms should be considered. The next relevant
term is quadratic in extrinsic curvature.
Its physical consequence is supposed  to make surfaces smooth.
This has been studied by both condensed matter physicists and string
theorists \cite{pfk,amp1}.
The analytical results obtained so far are the following. Perturbation
theory shows that the coupling of the extrinsic curvature term
is asymptotically free \cite{pfk,amp1}.
This is quite analogous to $2d$ classical
Heisenberg model, where according to Mermin-Wagner-Coleman theorem  there is
no long range correlation. It then appears that
a  physical length scale is generated through
dimensional transmutation, and in the infrared regime the extrinsic
curvature term is irrelevant and the model falls back to the Nambu-Goto
model \cite{amp1}. A large $D$ ($D$ being the embedding dimension) expansion
confirms this \cite{dg}.
On the other hand, numerous Monte-Carlo simulations \cite{cbr} of the same
system in small $D$ indicate that there is another fixed point,
so called crumpling transition point, away
from $\a=0$, that leads to scaling and vanishing of string tension at
the critical point. The transition may be second order. If so, a
sensible continuum limit may be defined.

It is the purpose of this note to develop  some qualitative understanding
of this new fixed point. Our strategy is the following. We seek for
a limit of the theory where an infrared fixed point exists. We then
extrapolate to the physical region and argue that the qualitative behavior
retains.
We first  assume decoupling of the Liouville mode in the
effective Lagrangian. The dynamical mechanism for this has been
proposed by Polchinski and Strominger \cite{joe}. We will apply their
result in our context.  We argue that the new form of the Liouville
action proposed in \cite{joe}
is also the only relevant term for our system.
The resulting effective action is exactly the one discussed in the
context of hexatic membrane \cite{dp}, except that in that case the coefficient
of the Liouville action can be arbitrarily large. Due to this difference,
we have to look for another approximation scheme.
There is no obvious expansion parameter other than $\a$.
It is known that one needs at least two parameters to predict cross-over
phenomenon. We find that small $D$ expansion is appropriate.
This is because 2-loop
contribution
can be comparable to that of 1-loop, while higher loops
can still be ignored.
The special role of the conformal anomaly is that it provides a relatively
large contribution with the desired sign
compared to 2-loop diagrams from the classical action.
This being the case, we can
 ignore those smaller contributions and the mathematics
becomes identical to that of hexatic membranes \cite{dp}.
It is then easy to see that a non-trivial fixed point indeed exists.
Unfortunately, small $D$ is not very physical -- it takes $D>2$ to have
the notion of extrinsic geometry. Our result can only be understood in terms
of analytical continuation. It
is meant to demonstrate that an alternative approach can lead to
new insight to the problem.
We can draw a renormalization  group flow diagram in
$(\a , D)$ plane. For small $D$, when $\a$ is large enough a flat phase
is present.
As $D$ becomes order of 1, two things can happen.
A phase transition of different nature than we are discussing here
may take place at $D=2$, as in the Polyakov string \cite{kpz}.
It seems that the critical indices are smooth at $D=2$, so we do not
expect the fixed point to disappear here.
Secondly our approximation can break down when
we extrapolate to $D\sim 3$. Our estimation shows that the 2-loop contribution
from classical action grows with $D$. When $D\sim 3$, it is comparable to
the Liouville contribution. The coupling constant
at the new fixed point is $\sim 1$, signaling the breakdown of perturbation
theory. Thus the approximation becomes very crude. Nevertheless, since
the numerical work shows a transition, our work can still point to a
possible albeit qualitative explanation.

The action for
a random surface with extrinsic curvature coupling can be written in
variety of ways \cite{amp,amp1}.
For our purpose, it is convenient to write it as
\begin{equation}
S=\int d^{2}\s \sqrt{g}( {\frac{1}{2\a}} \Box X_{\m}\Box X^{\m} + \tau ).
\end{equation}
The metric used in (1) is the induced metric
\begin{equation}
g_{ab}=\partial_{a}X_{\m}\partial_{b}X^{\m}.
\end{equation}
The specialty of the terms in (1) is that they are the only marginal or
relevant operators one can write down classically in physical space $R^{D}$.
That is, they behave like $\l^{n}$ ($n\ge 0$) under scaling $X^{\m}
\rightarrow \l X^{\m}$, thus are important at long distance.

We can of course study (1) by expanding around a fixed background. There is
however a question of functional measure. The standard way is to introduce
an intrinsic metric $g_{ab}$ and rewrite (1)  as
\begin{equation}
S=\int d^{2}\s \sqrt{g}[ {\frac{1}{2\a}} \Box X_{\m}\Box X^{\m} + \tau
+\l^{ab}(\partial_{a}X_{\m}\partial_{b}X^{\m}-g_{\a\b})].
\end{equation}
The new action introduces, among other things, new degrees of freedom,
in particular the conformal factor of $g_{ab}$.
The functional measure we use to evaluate the path integral of (4) is
the Polyakov measure \cite{amp2}.
As usual we choose the conformal gauge,
\begin{equation}
g_{ab}=e^{\phi} \hat{g}_{ab},
\end{equation}
where $\hat{g}_{ab}$ is a background metric which we take to be $\d_{ab}$
subsequently.
Following DDK \cite{ddk}, we would like to change
the functional measure from (l) to the one based on the background  metric
only. According to Polchinski and Strominger \cite{joe},
the Jacobian induced by the change
is not limited by the usual Liouville action, instead additional terms are
allowed. $\phi$ then becomes massive, and can be integrated out.
The action we end up with is
\begin{eqnarray}
S &=& \int d^{2}\s [ {\frac{1}{\a}} {\frac{1}{\partial_{z}X\cdot
\partial_{\bar{z}}X}}
\Box X_{\m}\Box X^{\m} + {\frac{\tau}{2}}\partial_{z}X\cdot\partial_{\bar{z}}X
+\l_{\bar{z}\bar{z}}\partial_{z}X_{\m}\partial_{z}X^{\m}+
\l_{zz}\partial_{\bar{z}}X_{\m}\partial_{\bar{z}}X^{\m}\, , \nonumber \\
&+&
{\frac{26-D}{48\pi}}{\frac{\partial_{z}(\partial_{z}X\cdot\partial_{\bar{z}}X)
\partial_{\bar{z}}(\partial_{z}X\cdot\partial_{\bar{z}}X)}{(\partial{z}X
\cdot\partial_{\bar{z}}X)^{2}}}].
\end{eqnarray}
One can easily show that the Liouville term in (5) is the only non-trivial
 term that
is renormalizable and invariant under $X\rightarrow \l X$, besides those
already appear in (1). All other terms with the similar property vanishes
in the gauge $\partial_{z}X\cdot\partial_{z}X=\partial_{\bar{z}}\cdot
\partial_{\bar{z}}X=0$.
There are not more relevant terms because they would have been seen in the
analysis of Polchinski and Strominger. There are of course terms that scale
with negative weight, they are irrelevant.
The measure used in (5) is simply the flat space measure.

The action (5) is power-counting renormalizable. To prove renormalizability
is however a non-trivial issue. One problem is  that $\partial_{a}X^{\m}$ is
dimensionless, thus any polynomial of it does not violate power-counting
renormalizability. This may not be serious because the scaling
invariance is only softly broken.
Another issue is that the Liouville term should
remain unrenormalized, since it comes from anomaly which is supposed
to be a 1-loop effect. Taking into account these, we see that essentially
there are only two constants that are renormalized, $\a$ and $\t$.
Here we assume that (5) is renormalizability.

Dynamical properties of (5) depend on both $\a$ and $\t$. It is useful
to think (5) as a Landau theory of phase transition, where $\t$ plays the
role of $T-T_{cl}$. When $\t$ is large, (5) reduces to the Nambu-Goto
model. When $\t$ is small, infrared divergence is present and renormalization
group (RG) approach is necessary. We are mainly interested in the RG behavior
of $\a$.
We  first compute 1-loop diagrams without the Liouville contribution.
It is a repetition of a well-known result \cite{pfk,amp1}.
Since our gauge choice and
the path integral measure differ from the previous works, it
is worth doing the calculation again.
As usual background field method is useful. We expand $X$ as
$X^{\m}=\bar{X}^{\m}+Y^{\m}$, where $Y^{\m}$ is quantum fluctuations and
$\bar{X}^{\m}$ is non-static background.
It is more convenient to use Cartesian coordinates. The action (8) is rewritten
as
\begin{eqnarray}
S &=& \int d^{2}\s [ {\frac{1}{\a}} {\frac{1}{\partial_{a}X\cdot
\partial^{a}X}}
\Box X_{\m}\Box X^{\m} + {\frac{\tau}{2}}\partial_{a}X\cdot\partial^{a}X \,
\nonumber \\
&+&
{\frac{26-D}{48\pi}}{\frac{\partial_{a}(\partial_{b}X\cdot\partial^{b}X)
\partial^{a}(\partial_{c}X\cdot\partial^{c}X)}{(\partial{d}X
\cdot\partial_{d}X)^{2}}}].
\end{eqnarray}
To quadratic order in the quantum field $Y^{\m}$, the action without
the Liouville contribution is
\begin{eqnarray}
S_{Q} &=&{\frac{1}{\a}}
 \int d^{2}\s [{\frac{1}{\partial_{a}\bar{X}\cdot \partial^{a}\bar{X}}}
\Box Y \cdot \Box Y -4{\frac{\partial_{b}\bar{X}\cdot \partial^{b}Y}
{(\partial_{a}\bar{X}\cdot \partial^{a}\bar{X})^{2}}}\Box \bar{X} \cdot \Box Y
\, \nonumber \\
&+& {\frac{\Box \bar{X}\cdot \Box \bar{X}}{\partial_{a}\bar{X}\cdot
\partial^{a}\bar{X}}}(-{\frac{\partial_{b}Y\cdot\partial^{b}Y}
{\partial_{a}\bar{X}\cdot \partial^{a}\bar{X}}}
+ 4 {\frac{(\partial_{b}\bar{X}\cdot\partial^{b}Y)^{2}}
{(\partial_{a}\bar{X}\cdot \partial^{a}\bar{X})^{2}}})
+{\frac{\t}{2}}\partial_{a}Y\cdot\partial^{a}Y].
\end{eqnarray}
In order to do perturbation theory, we expand the background  $\bar{X}^{\m}$
around flat space,
\begin{equation}
\partial_{a}\bar{X}^{\m}=e_{a}^{\m}+\partial_{a}\tilde{X}^{\m},
\end{equation}
here $e_{a}^{\m}$ is constant and the last term is a small fluctuation.
We then do perturbation theory with respect to $\partial_{a}\tilde{X}^{\m}$.
Propagator is
\begin{equation}
Y^{\m}(p)Y^{\n}(-p) = {\frac{\a {e\cdot e}}{2}}
\d^{\m\n}  {\frac{1}{p^{4}+\t p^{2}}},
\end{equation}
where we have defined  $e\cdot e \equiv e_{a}^{\m}e_{a}^{\m}$.
Before we do loop calculation, let's comment on the regularization.
We want to cut-off both long and short wave-length contributions.
The cut-offs should however be in real space, not on world sheet.
If we denote the momentum  cut-offs in the embedding space
as $\L_{max}$ and $\L_{min}$, for high and low momenta respectively,
then the integration region  of world sheet momenta are
$\sqrt{e\cdot e}\L_{\max}$ and $\sqrt{e\cdot e}\L_{min}$, respectively.
If we keep $\t \ne 0$, $\sqrt{e\cdot e} \L_{min}$ is replaced by
$\sqrt{e\cdot e} \t$.
There are two diagrams contributing to the 1-loop $\b$ function.
The tadpole contribution from the last term in (7) is
\begin{equation}
-{\frac{D-2}{4\pi}} \ln ({\L_{max}/\t}) \int d^{2}\s \, {\frac{1}{e\cdot e}}
\Box \tilde{X}\cdot \Box \tilde{X},
\end{equation}
while the other diagram, constructed from from the second term in (7),
gives
\begin{equation}
-{\frac{1}{2\pi}} \ln ({\L_{max}/\t})\int d^{2}\s\,
 {\frac{1}{e\cdot e}} \Box \tilde{X}\cdot
\Box \tilde{X}.
\end{equation}
The sum is
\begin{equation}
-{\frac{D}{4\pi}} \ln ({\L_{max}/\t})\int d^{2}\s\,  {\frac{1}{e\cdot e}}
\Box \tilde{X}\cdot \Box \tilde{X}.
\end{equation}
We define renormalized $\a_{r}$  at scale $\m$ as
\begin{equation}
{\frac{1}{\a_{r}}} = {\frac{1}{\a_{0}}}- {\frac{D}{4\pi}}
 \ln (\L_{max}/\m).
\end{equation}
The $\b$ function is then \cite{pfk,amp1}
\begin{equation}
\b(\a)={\frac{d\a}{d\ln\m }}=-{\frac{D\a^{2}}{4\pi}}.
\end{equation}
Next order contribution to $\b$ will be $O(\a^{3})$. So if $D$ is small,
a $D$ expansion is viable. Now
we  have two small parameters, $D$ and $\a$. Non-trivial
renormalization group flow can exist as the relative strength changes.
2-loop effects are now very important. As we will argue later that the $O(\a)$
correction to (13) from the classical action is small when $D$ is small, the
main contribution is from the Liouville action, due to the
largeness of 26. Although the diagrams are
1-loop, they are actually 2-loop effect.
Once we ignore other 2-loop contributions, we are dealing with
the same problem as the hexatic membrane \cite{dp}, but the assumption is
very different.

Now we expand the Liouville action to the second order in $Y$,
\begin{eqnarray}
S_{L} &=& {\frac{K}{8}}\int d^{2}\s
\lbrace 4\partial_{b}({\frac{\partial_{c}\bar{X}\cdot \partial^{c}Y}{
\partial_{a}\bar{X}\cdot \partial^{a}\bar{X}}})
\partial^{b}({\frac{\partial_{d}\bar{X}\cdot \partial^{d}Y}{
\partial_{e}\bar{X}\cdot \partial^{e}\bar{X}}})\, \nonumber \\
&-& 2\Box \ln (\partial_{a}\bar{X}\cdot \partial^{a}\bar{X})[{\frac
{\partial_{b}Y\cdot\partial^{b}Y}{\partial_{c}\bar{X}\cdot
\partial^{c}\bar{X}}}
-2({\frac{\partial_{d}\bar{X}\cdot\partial^{d}Y}
{\partial_{e}\bar{X}\cdot \partial^{e}\bar{X}}})^{2}]\rbrace,
\end{eqnarray}
where $K\equiv (26-D)/12\pi$.
The first term in (15) contributes to the $\beta$ function.
The other terms may give rise divergence to the Liouville
action. We verified that this is not the case.
We now further expand the background in (15).
Writing out relevant terms, we find
\begin{eqnarray}
S_{L} &=& {\frac{K}{2}}
\int d^{2}\s [{\frac{\partial_{a}\partial_{b}\tilde{X}\cdot
\partial_{b}Y}{e\cdot e}} {\frac{\partial^{a}\partial_{c}\tilde{X}\cdot
\partial_{c}Y}{e\cdot e}}-2{\frac{\partial_{b}\tilde{X}\cdot
\partial_{b}Y}{e\cdot e}}{\frac{\Box\partial^{c}(e_{c}\cdot Y)}{e\cdot e}}\,
\nonumber \\
&+& {\frac{\partial_{a}(e_{b}\cdot Y)}{e\cdot e}}
{\frac{\partial^{a}(e_{b}\cdot Y)}{e\cdot e}}
+\cdots].
\end{eqnarray}
There are three kinds of contributions. One is tadpole from the first term
in (16), it gives
\begin{equation}
K{\frac{1}{8\pi}}(\L/\t) \ln ({\L_{max}/\t})\int d^{2}\s\, {\frac{1}{e\cdot e}}
\Box \tilde{X}\cdot \Box \tilde{X}.
\end{equation}
The diagram involving the second term in (16) does not
 contain logarithmic divergence.
Other diagrams involves one insertion of the last term of (16) into the two
diagrams involving classical vertices of (7).
We can insert it to the tadpole of classical action, it gives
\begin{equation}
K{\frac{1}{8\pi}}\ln ({\L_{max}/\t})\int d^{2}\s\,  {\frac{1}{e\cdot e}}
 \Box \tilde{X}\cdot \Box \tilde{X}.
\end{equation}
We can also insert it to the diagram with the two classical vertices, its
contribution is
\begin{equation}
 -K{\frac{1}{16\pi}}\ln ({\L_{max}/\t})\int d^{2}\s\,  {\frac{1}{e\cdot e}}
 \Box \tilde{X}\cdot \Box \tilde{X}.
\end{equation}
Summing over (17)-(19), we have
\begin{equation}
K{\frac{3}{16\pi}}\ln ({\L_{max}/\t})\int d^{2}\s\,  {\frac{1}{e\cdot e}}
 \Box \tilde{X}\cdot \Box \tilde{X},
\end{equation}
which agrees with \cite{dp}.

The $\b$ function to this order is
\begin{equation}
\b(\a)=\a^{2}(-{\frac{D}{4\pi}}+{\frac{3K\a}{16\pi}}).
\end{equation}
There is a infra-red fixed point at
\begin{equation}
\a^{*}={\frac{4D}{3K}}.
\end{equation}
As long as $D$ is small, we have a consistent expansion. At the fixed point,
the rigidity $1/\a^{*}$ is large but finite, and the string tension vanishes.
Near the fixed point, the string tension may scale. Such a transition is
called crumpling transition.

Let us draw a possible
phase diagram in $(\a, D)$ plane.
When $D$ is small, there exists a infrared fixed point at some $\a^{*}$.
So there is a fixed line in the $(\a,D)$ plane.
The line is infrared attractive.
The situation at large $D$ is less clear. It is possible that the curve
bends over at some $D^{*}$ so that there is no infrared  fixed point
at large $D$ (Fig.1a). On the other hand the curve may persist in large $D$
(Fig.1b).
There is no contradiction between this behavior and the $1/D$ expansion,
since there one assumes $\a\sim 1/D$.
In the following we will make an estimate from  small $D$ approximation.
Let us first compare the 2-loop contribution from the classical
action with that of the Liouville action (20).
We must expand (6) up to quartic order in $Y$. Since it is
not very illuminating, we will not write it down here.
What we can do is to note that the effective coupling constant of (6) is
$\a/4\pi$, where a $2\pi$ factor is from loop integration and another 2
is seen from the propagator.
Using this rule we estimate the 1-loop contribution is roughly $1/4\pi$
(see (13)). The Liouville contribution is $\sim K/4\pi$ (see (20)).
They are roughly correct. So the 2-loop contribution is similarly
guessed as $\a (D+c)/16\pi^{2}$, where $c\sim \pm 1$.
Compare this with (20) we see that they are of the same magnitude
when $D\sim 3$.
Depending on the constant piece, the 2-loop result may be smaller than  (20).
Of course $\a^{*}/4\pi$ is of order $1$ when $D$ is around 3. So numerically
we do not have a good approximation in physical dimensions.
However qualitatively the diagram should be correct.

Another question is whether the curve bends over when $D=2$. $D=2$
is special because (1) has physical degrees of freedom when $D\ge 2$ and
it is a critical point of the Polyakov string \cite{kpz}.
Here if we use naively
the results of \cite{dp} on various critical indices around
the fixed point, we find that although the signs change when we cross $D=2$,
no singularity exists. This means that the change at $D=2$ is milder
than in Polyakov string. It is likely that the $D=2$ barrier is
overcome by introducing the extrinsic curvature coupling.

So far we have not discussed the $RG$ flow of the string tension $\t$ around
$\t\sim 0$.
According to \cite{pfk,dp}, the anomalous dimension of the string tension
is proportional to $-(D-2)$.
When $D>2$,  it is infrared unstable and the fixed point $\a^{*}$
is a critical point.
When $D>2$, the fixed point is stable so that it does not correspond to
critical point.
On the other hand, Polyakov \cite{amp1} pointed out
that due to asymptotic freedom, a non-zero string tension $\t^{*}$ may be
generated by dimensional transmutation, i.e., there is a stable flow
towards $\t^{*}$. Combine both $\a$ and $\t$ flows together, we have
conjectured the flow diagrams for $D>2$ (Fig.2a or 2b)
 and $D<2$ (Fig.3), respectively.

If these guesses  are true, we conclude that for $D=3,4$ the critical point
exists and a flat phase is realized.
This is in agreement with the numerical
simulations \cite{cbr}. Or rather, the numerical work prompts us to believe
that
the conformal anomaly has something to do with the appearance
of nontrivial fixed point of $\a$. The application
of (1) in particle physics
 is however not clear: originally it was thought that the continuum
limit of (1) might describe $QCD$ string. The recent work of Polchinski
\cite{joe2} seems to suggest that such a expectation may not be realized
in $N_{c}\rightarrow \infty$.

As a final remark, we point out that it is not clear whether  physics at
the critical point is described by a conformal field theory. The reason
is that scale invariance is equivalent to conformal invariance only
when very specific conditions are met \cite{zp}. The present model
is a higher derivative theory, so unitarity in Minkowski space
is not obvious. Without unitarity, the equivalence proof does not go through.
Parenthetically we remark that classically
the scale invariance is not broken when
we expand around a constant
classical background $\partial_{a}X^{\m}_{cl}$, because  being dimension zero
the background does not scale.

\end{document}